\documentstyle[12pt,epsf,epsfig,rotating]{article}
\begin{document}
\noindent

\title{\bf Synchrotron Polarization \\ at \\ High Galactic Latitude}

\author{Z. Zainal Abidin$^{1}$\footnote{email: {\tt oezzai@jb.man.ac.uk}},
 J. P. Leahy$^{1}$, A. Wilkinson$^{1}$, P. Reich$^{2}$, W. Reich$^{2}$ \\
and R. Wielebinski$^{2}$
\\\small{\it $^{1}$University of Manchester, Jodrell Bank Observatory,
SK11 9DL Cheshire, England}\\
\small{\it $^{2}$Max-Planck-Institut f$\ddot{u}$r Radioastronomie, 
Postfach 2024, 53010 Bonn, Germany}}

\date{\footnotesize{\today}}

\maketitle

\abstract{We present preliminary results from mapping the
high-latitude Galactic polarization with the Effelsberg Telescope at
$\lambda$21 cm. Structures on the resulting maps are mostly on the
scale of several degrees. The results show detection of polarization
over most of the field, at the level of tens of percent of the
synchrotron emission. The evidence of more structure in Stokes $Q$
and $U$ rather than in $\sqrt{Q^{2}+U^{2}}$ suggests the existence of
Faraday rotation.}

\section{Introduction}

If temperature anisotropies are results of primordial fluctuations,
then their scattering at the last scattering surface would be
imprinted in a polarization signal. Verification of the Cosmic
Microwave Background (CMB) polarization would thus give us clearer
picture of how primordial fluctuations behave. CMB polarization,
together with CMB temperature anisotropies are important in the study
of the acoustic peaks in the CMB angular power spectrum. CMB
polarization itself is specifically used in studying the general
reionization period and  the B-mode
polarization is a possible indicator of gravitational waves. 
The dominant problem in observing the CMB polarization
is foreground emission.

The aim of this work is to understand more clearly the most polarized
foreground of them all, which is the synchrotron emission,
specifically at high Galactic latitude. At this latitude, the
synchrotron polarization dominates the foreground up to 100
GHz \cite{2002apb..conf..109C}.  Because of the much shorter path through
the Galactic disk compared to the better-observed lower latitudes,
fixed angular scale corresponds to a smaller physical
scale size, and also the amount of structure along any
given line of sight should be much smaller; therefore we cannot simply
extrapolate to high latitudes the polarization structure seen near the 
Galactic plane. 

The seperation of the CMB and synchrotron signals is made by combining
observations of the same area of sky made at different
frequencies.  Although it has been suggested that this extraction can also be 
made from single frequency data, even when the polarized CMB signal is just a 
fraction of the total polarized signal, through a statistical
estimator \cite{2002astro.ph..9400S}, this requires assumptions
about the statistical structure of the foreground emission, e.g. a pure
power law power spectrum. But since there is every reason to expect that such 
assumptions are at best poor approximations, and since there is absolutely
no way to check their accuracy except by multi-frequency observations, 
such observations really are essential.

\section{Synchrotron Polarization and Faraday Depolarization}

In astrophysics, much of the observed radio frequency emission from
supernova remnants, active galactic nuclei and numerous other sources
are in the form of synchrotron radiation.
Its polarization ranges from around 10\% to up to 75\% and, as mentioned
before, is the dominant foreground up to a frequency of about 100~GHz. 
Because of its steep spectrum, observations of the diffuse synchrotron 
polarization are usually at low frequencies; for instance at high Galactic
latitudes, the Dwingeloo survey at 1.4~GHz, conducted more than 30 years ago
\cite{1976A&AS...26..129B} provides the highest frequency for which significant areas
of the sky are covered.

Unfortunately, depolarization from Faraday rotation becomes important at long
wavelengths, and on the basis of the rotation measures of extragalactic 
sources, detectable rotation is expected all over the sky at 
$\lambda \ge 20$ cm. In contrast, CMB measurements are in the 
Faraday-thin regime.  How will the distribution of observed polarization
be affected by this difference?  The effects of Faraday rotation have 
been studied both theoretically \cite{1979MNRAS.186..519B, 1991MNRAS.250..198G, 1991MNRAS.250..726T} and by
direct multi-frequency observations \cite{2003MNRAS.342..399G} and the following points
seem generic:
\begin{itemize}
\item As the observing wavelength increases, the initial effect is rotation, 
not depolarization, even when the Faraday-rotating and emitting regions
are mixed, as they must be to at least some extent for the Galactic 
polarization. Although this has no effect on the polarized intensity, it
moves flux from Stokes $Q$ to $U$ and vice versa. In general this moves
power to higher spatial frequencies. 
\item One might imagine that the imposed angular scale is that of the 
magnetic field responsible for the Faraday effect, but this is not so. 
Once the rotation angles become large ($> 1$ rad), $Q$ and $U$ 
(or $E$ and $B$) oscillate on the scale where the differential rotation is 
$\pi$ rad, i.e. 
\[ \ell \approx 2 |\nabla_\theta RM| \lambda^2. \]
That is, even smooth gradients in the $RM$ suffice to move structure to
high $\ell$ at long wavelengths.
\item At least in the extragalactic case, the $RM$ distribution is not a 
smooth field (even though $RM$ is a scalar) but contains 
discontinuities, visible both as jumps in the $RM$ (where multi-frequency
data is available), and as lines of strong depolarization due to large differential rotation within the beam. By construction,
these correspond to current sheets in the Faraday-active medium seen edge
on, suggesting that the current in the plasma is quite intermittent, as
expected from MHD simulations \cite{2000PhPl....7.4889B}. Similar depolarized filaments
are seen in the Galactic emission, particularly in the
data of Uyan{\i}ker et al. \cite{1999A&AS..138...31U}; it is controversial whether these represent
the same phenomenon since so far the $RM$ distribution is not clearly mapped.
\item At somewhat longer wavelengths the polarization is is wiped 
out altogether by differential rotation, both along the line of 
sight, and transverse to the line of sight within the beam. 
According to the simplest model of an infinitely tangled field \cite{1966MNRAS.133...67B}
the depolarization is very rapid; the degree of polarization behaves as
$m(\lambda) = m_0 \exp[-\sigma_{RM}^2 \lambda^4]$, almost a step function 
in terms of log(frequency).
In practice the polarization of extragalactic sources falls much more slowly
than this, more like a power law. High resolution imaging (e.g. 
\cite{1995MNRAS.273..877J}) shows that this
is mainly due to large variations in $\sigma_{RM}$ between different
regions, and depolarization is indeed rapid when the structure in the $RM$
screen is resolved.
\end{itemize}

Clearly, attempts to extrapolate the polarized power spectra observed
in the Faraday-thick regime to the Faraday-thin regime are doomed.  At
present, we do have large-area surveys at $\lambda$13 cm
\cite{1997MNRAS.291..279D} and $\lambda$11 cm \cite{1987A&AS...69..451J}, 
but these are along the Galactic plane where the $RM$ of
extragalactic sources is typically 1-2 orders of magnitude larger than
at high latitudes. Inspection of these surveys shows less polarized
flux (not just fractional polarization) along the plane compared to
the edges of the survey at slightly higher Galactic latitudes.  Along
the long lines of sight through the plane, we expect to encounter many
different intrinsic magnetic field directions, and so the polarization
can be considered the result of a vector random walk in the $(Q,U)$
plane.  This will result in a lower degree of polarization than at high
latitudes (since the total intensity adds linearly), but nevertheless
the polarized flux should be {\em higher} because of the longer random
walk.  Therefore the low polarization along the plane at $\lambda 11$~cm 
must be an effect of Faraday rotation, not just line-of-sight
averaging. This is entirely as expected, given data on extragalactic
sources.  Evidently, even these results cannot be used to predict the
foreground in the CMB wavelength window.

To correct CMB data we need accurate maps of the synchrotron polarization
at medium and high latitudes. While it may be possible to extract this
directly from the {\it WMAP} or {\it Planck} multi-frequency data, this
could be confused if there is another polarized component (e.g. spinning
dust), and in any cases it is safest to have a good template at a frequency
where the synchrotron component dominates. At high latitudes, typical 
rotations of $\sim$0.5 rad are expected at 1.4~GHz; these are
in the intermediate regime where depolarization is relatively small, so
could be corrected with multi-frequency data, or, ideally, by observing at a 
substantially higher frequency.  Unfortunately, large-area surveys with
existing equipment in the 1--10 GHz band have difficulty in measuring the 
absolute polarization because of baseline drifts (essentially, $1/f$ noise),
as discussed below.

This motivated our test survey to see if absolute polarization could be
obtained with a different mapping strategy. 

\section{Mapping Strategy}

The survey is done with the Effelsberg 100-m single dish telescope. It
has a beamwidth of $9'\!.35$ at $\lambda$21 cm (1.4 GHz). The region of the 
survey is shown in Fig. 1. It passes just north of the North Galactic Pole,
and so of the North Polar Spur, and at its western end 
contains the coldest part of the sky at 408 MHz. 
The two filter centre frequencies are at 1395 MHz and 1408~MHz. 

Fig. 1 also shows a rough idea of the scanning strategy. The observation
was done at a constant elevation of 55$^{\circ}$ in order to minimise
variation in spurious polarization, i.e. elevation-dependant signal in
the far sidelobes. Individual sub-scans were $18^\circ$ long, and we
scanned at the fastest possible speed ($8^\circ$/minute) to minimize system
fluctuations during each scan (for our observations this was limited by 
gearbox problems; faster scanning is now possible). Calibration observations
were made at the normal speed of $4^\circ$/minute.

Scans are done on both the rising and setting side of
the survey declination strip. This give us crossed scans, allowing
`self-calibration'. A fully-sampled survey would required 12 nights on
each side. In practice, we obtained measurements on 6 nights on the
East and then 5 nights on the West causing undersampling at the edge
of the strip. However, the sampling is twice as dense in the centre
(Fig. 1) and so we are essentially fully sampled there. The survey
is done at night to eliminate solar interference. A significant disadvantage
of this approach is the requirement to start each night's observation at a 
precise siderial time, with the associated problem that any brief
interruption can put many sub-scans out of place.

\begin{figure}
\begin{center}
\includegraphics[width=5.0cm, angle=-90]{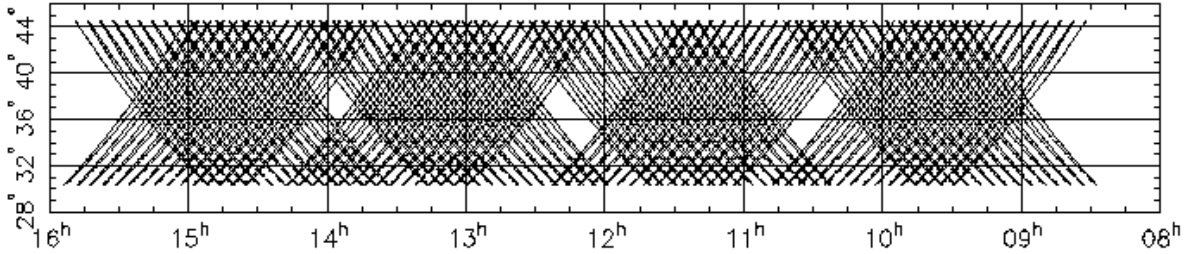}
\caption{\label{out2} Illustration of the Mapping strategy. The zig-zag
lines show an idealised version of the scan pattern from a single night's
observation on each of the east and west sides. Gaps in the pattern
on each night occur when we made calibration scans of the North pole,
and the larger gap includes time out to observe the primary calibrator,
3C\,286. Observations on other days are slightly shifted to build up a
well-sampled sky image.}
\end{center}
\end{figure}

The length of sub-scan (i.e. continuous track in one direction) 
was a compromise: longer sub-scans are more efficient, as
time spent reversing direction at the end is minimized,
but they require more interleaved days to build up complete sampling.
If sufficient telescope time were available for a survey of a significant 
fraction of the sky, longer scans would be preferable.

\section{Survey Calibration and Systematics}

One particular calibration technique for a sensitive 1.4 GHz continuum
and polarization survey with the Effelsberg 100-m telescope is
described by Uyan{\i}ker et al. \cite{1998A&AS..132..401U}. Their survey
observed areas up to $b = \pm$20$^{\circ}$ with the same system used by us.
The survey was done in $10^\circ$ square patches, scanning
in Galactic latitude and longitude. This is a much more
straightforward approach than ours, allowing scanning of contiguous regions
in a single night and not subject to problems of synchronization from
night to night. The drawback is that elevation-dependent artefacts occur,
which must be removed by subtracting a low-order polynomial from each scan.
This also takes care of instrumental drifts, but removes all astronomical 
structure on scales larger than a few degrees. Uyan{\i}ker et al. were able
to restore the missing large-scale structure using information from 
low-resolution surveys, in particular the Dwingeloo 1.4 GHz measurements 
for polarization.  However, the area covered by Dwingeloo did not coincide 
with our survey, and in any case our aim was to test a method which
could be used at other frequencies, where no prior data is available.

In our survey we followed standard practice in most respects. 
Point source calibrators (usually 3C\,286) were used to determined the 
gains correction factors, instrumental polarization, 
and the absolute polarization position angle. The data reduction used
the NOD2 package \cite{1974A&AS...15..333H}.  

With elevation effects eliminated by our observing strategy, we were able to
capitalize on the excellent receiver stability at Effelsberg
to determine polarization zero levels absolutely, using observations of the 
North Celestial Pole (NCP) every 1--2 hours. 
The true polarization of the pole was determined
via its rotation with parallactic angle: separate estimates for each 
night, initially assuming constant offsets, were averaged. 
Using the average NCP polarized emission, 
the slow variation of the offsets during each 
night could then be tracked. This offset is 
believed to be due mainly to receiver noise leakage throught the OMT, 
which allows a correlated signal; typical values are $\sim 1$~K, 
$\sim 4$\% of the system temperature.

We also used the NCP to track the total 
power offset, up to  an arbitrary zero level. Offsets (and hence gains) were 
stable to within a few $\%$ on each night, but in total power
the variation was still large compared to the structure in the sky emission.  
We initially fitted the NCP offsets with
a low-order polynomial, and subtracted this from the survey data. 
While this still left clear differential offsets between observations on
different nights, these were smaller than the overall brightness trend along
the survey strip. We further reduced the residual offsets by averaging the
modal (most common) pixel levels for each survey field over the 5--6 sets of interleaved data,
and then setting the modes of each dataset equal to the field average. 
In total intensity only, scanning effects were also reduced by the usage 
of unsharp masking \cite{1979A&AS...38..251S}. Further 
destriping, using the crossed scans from East and West side observations, will
be applied in the future. 

The total offset includes stray radiation (or spillover) which is 
dependent on the pointing direction, in principle including azimuth due
to local topography. In fact, despite the telescope's location in a
narrow valley, we found no sign of an azimuth effect, except that
one half of the survey suffered from weak interference, giving a 
fixed-azimuth pattern that was modelled and removed for each night. 

The small difference between the elevation of the NCP 
($\mu = 50^{\circ}\!.25$) and of our survey scans causes a
negligible differential polarized spillover, as confirmed both by test
observations which showed that the average level of the polarization 
signal did not vary detectably with elevation in the range 
$50^{\circ} \leq \mu \leq 60^{\circ}$,
and by the fact that the polarized flux in part of our survey known to
have minimal synchrotron emission is close to zero after subtracting offsets
as detailed above (see Section 5). There is a significant (but constant) 
effect in total power, giving an overall negative offset to our images.

\section{Results}
\label{results}

Figs. 2 and 3 show the result of a single field in total power and
polarization respectively.

\begin{figure}
\begin{center}
\includegraphics[width=9.0cm, angle=-90]{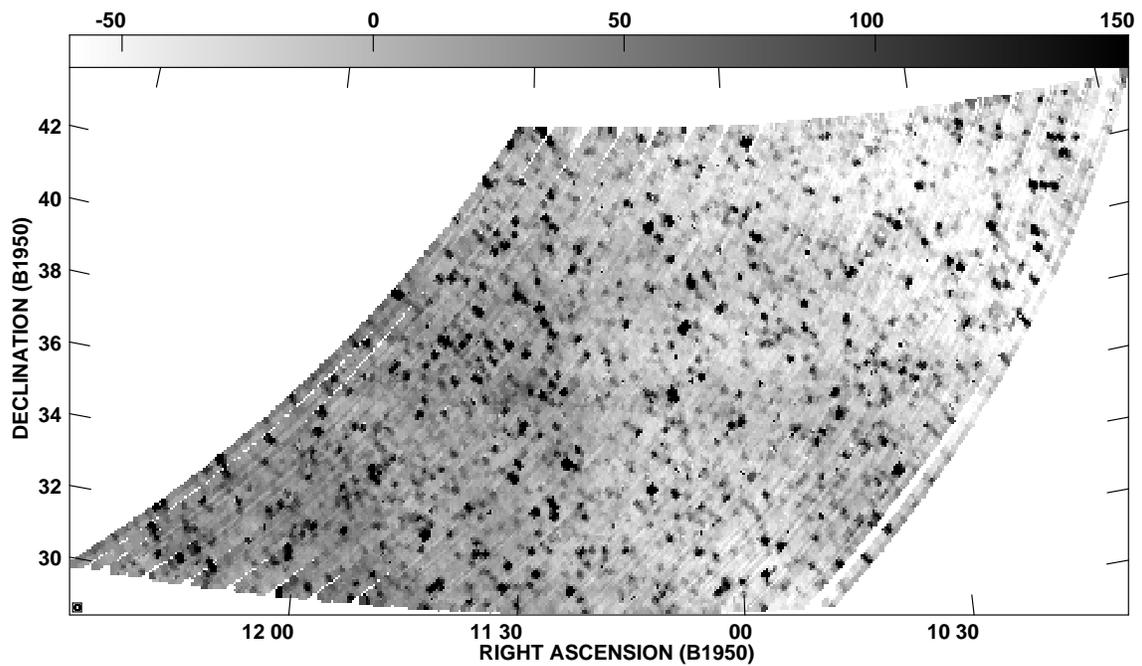}
\caption{\label{out3} Total Power Map: Units are mJy beam$^{-1} = 2.12$~mK.
Data has been interpolated onto a course grid to alleviate the effects 
of undersampling, but this is still apparent at the north and south edges 
of the survey. The zero level is arbitrary.}
\end{center}
\end{figure}

\begin{figure}
\begin{center}
\includegraphics[width=9.0cm, angle=-90]{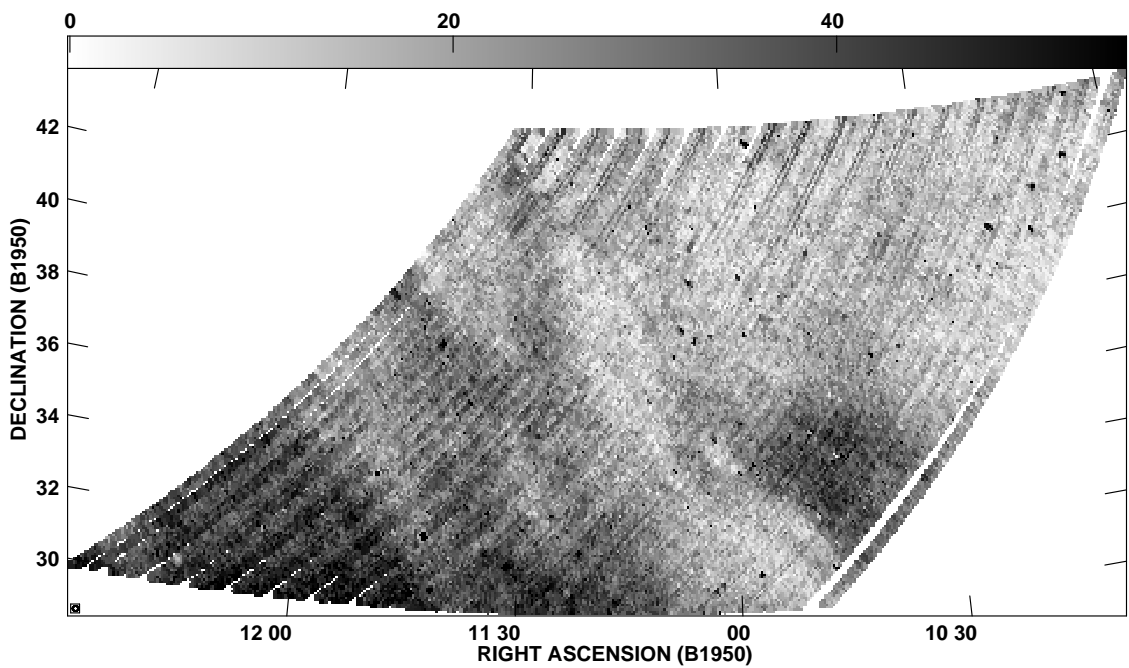}
\end{center}
\caption{\label{out4} Map of polarized intensity, with close to absolute
zero level. Units as in Fig. 2.}
\end{figure}

The results shows detection of polarization over most of the
field. Structures are mostly on the scale of several degrees.
In the cold region at the western end of our survey strip 
our corrected polarized flux is close to zero. On this basis
we believe that our the polarization zero levels are correct to within
around 8 mK (2$\sigma$). 

The mean polarized intensity over the whole strip is about 70 mK,
while the polarized intensity peaks at about 130 mK. The root mean
squared (RMS) of the $Q$ and $U$ signal is calculated around 50 mK, after 
subtracting an estimate of the noise in quadrature. Taking the `cold' region
as close to zero synchrotron emissivity, a very rough estimate for the
fractional polarization of the diffuse emission is 30-40\% in the brighter 
regions.  In contrast, background extragalactic sources are typically 
a few percent polarized or less, so that while the total power images are
limited by confusion from point sources, only a few of these show up in 
polarization.

There is perceptibly more structure in $Q$ and $U$ than in polarized 
intensity, $p = \sqrt{Q^{2}+U^{2}}$, indicating that position angle changes 
on a shorter length scale than $p$.  We would expect synchrotron emission
to give structure in angle and intensity on roughly the same scale, so this
may suggest that even at high latitudes 1.4 GHz polarization is affected by
Faraday rotation. This is expected on the basis of rotation measures of
extragalactic sources, $\sim 10$ rad~m$^{-2}$ at high Galactic latitudes,
corresponding to rotations of $\sim 20^{\circ}$ at our frequency. However, this
is not enough to cause significant depolarization, especially as the
angular scale of position angle variation 
is much larger than the resolution of the observations, so no beam
depolarization is likely.

\section{Outlook}

Final calibration and destriping is in progress and it is expected to
be followed by a detailed analysis of the power spectrum provided by
this survey.

The suggestion of the Faraday rotation effect in the results so far
requires checking. A suitable baseline in frequency, given the expected
rotation measures, would be provided by a second survey at $\lambda$18 cm
which could be made with the same system and could be matched in resolution
with minimal smoothing.  Where signal-to-noise is high, this would allow
extrapolation of the position angle to zero wavelength, appropriate for
comparison with data from the CMB bands.  A better solution
would be a survey at short enough wavelength that Faraday rotation could be
neglected; at high latitude, $\lambda$6 cm would suffice. A 15-25 m dish 
would give similar resolution to our work (and to {\it Planck}).

By itself, this survey will provide a better estimate of the statistical
properties of the polarization at high latitude, and as such will facilitate
the  study of detailed foreground contamination of the polarized CMB.

\bibliographystyle{unsrt}
\bibliography{ref}

\end{document}